\documentstyle[prb,aps,amssymb,graphicx]{revtex}

\begin{document}

\draft
\twocolumn[\hsize\textwidth\columnwidth\hsize\csname @twocolumnfalse\endcsname

\title{Fractional and Integer Excitations in Quantum
Antiferromagnetic Spin $1/2$ Ladders}

\author{C. Knetter$^1$,
K.P. Schmidt$^1$, M. Gr\"uninger$^2$, and G.S.~Uhrig$^1$}
\address{$^1$Institut f\"ur Theoretische Physik, Universit\"at zu
  K\"oln, Z\"ulpicher Stra\ss{}e 77, D-50937 K\"oln, Germany}
\address{$^2$II. Physikalisches Institut, Universit\"at zu
  K\"oln, Z\"ulpicher Stra\ss{}e 77, D-50937 K\"oln, Germany}

\date{\today}

\maketitle
\begin{abstract}
Spectral densities are computed  in unprecedented detail for
quantum antiferromagnetic spin $1/2$ two-leg ladders. These
results were obtained due to a major methodical advance achieved
by optimally chosen unitary transformations. The approach is based
on dressed integer excitations. Considerable weight is found at
high energies in the two-particle sector. Precursors of fractional
spinon physics occur supporting the conclusion that there is no
necessity to resort to fractional excitations in order to describe
features at higher energies.
\end{abstract}



]

Low-dimensional quantum antiferromagnets are of fundamental and
enduring interest. One reason is that high-temperature
superconductivity depends crucially on the interplay of charge
carriers with the magnetic excitations of a two-dimensional (2D)
quantum antiferromagnet. A vivid debate concerns the nature of the
elementary excitations of such an antiferromagnet. In terms of
integer spin waves ($S=1$), no quantitative description is
available of the spectral densities at higher energies, where a
significant part of the spectral weight is located
\cite{aeppl99,gruni00}. Therefore, it has been suggested that
 fractional ($S=1/2$ spinons) excitations play an important role
in 2D \cite{ho01} supporting
the view that spinons and spin-charge separation are the basis of
 high-$T_c$ superconductivity \cite{laugh97,ander00}.
Spin ladders are similar to 2D planes in the range of higher energies.
The importance of fractional excitations for the description of high
energy excitations in 2D would hence imply their importance in spin
ladders. Here we present a description
of the spectral densities in unprecedented detail in
terms of {\em integer} excitations. Our results show that the
essential point is not the fractionality of the elementary
excitations (``particles'') but the proper description of
multi-particle excitations.

Spectral densities provide information on the density of elementary
excitations, on their interaction
and on how the particular excitation operator
couples to them. For instance the dynamic structure factor
as measured  by inelastic neutron scattering couples to  excitations with
total spin $S=1$. Integer spin excitations induce
a dominant so-called quasi-particle peak
in the dynamic structure factor \cite{sandv01}.
If the integer $S=1$ excitation decays into two
fractional $S=1/2$ spinons, which move independently at large distances,
 the quasi-particle peak vanishes and
an important continuum appears at higher energies. The generic example
is found in spin chains for which spinons are the elementary excitations
\cite{karba97}.

We address 2-leg spin ladders, i.e.\ two chains with
intra-chain coupling $J_\parallel$ coupled by an exchange coupling
$J_\perp$
\begin{equation}
\label{hamilton} H = \sum_{i}[J_\parallel({\bf S}_{1,i}{\bf
S}_{1,i+1} + {\bf S}_{2,i}{\bf S}_{2,i+1})+J_\perp {\bf
S}_{1,i}{\bf S}_{2,i}]\ ,
\end{equation}
where $i$ denotes the rung and $1,2$ the leg. As far as the
excitations at higher energies  are concerned, spin ladders
constitute an intermediate system between 1D chains and 2D square
lattices. Additionally, they
 occur in a multitude of substances
\cite{johns00a} comprising also superconducting systems
\cite{uehar96}.

What is the most effective way to describe the excitations?
 Two approaches are
possible: (i) If the rung  coupling $J_\perp$ dominates in
(\ref{hamilton}), i.e.\ $H_0:=H\big|_{J_\parallel=0}$ the
excitations are local triplets on each rung (rung-triplets). When
$J_\parallel$ is switched on they start to hop from rung to rung.
On increasing $J_\parallel$ the excitations become more and more
extended; they are rung-triplets dressed by a magnetically polarized
environment. (ii) If the rung coupling
$J_\perp$ is weak it can be treated as perturbation which acts on
the elementary spinons in both chains. The rung coupling binds two
spinons on either chain thereby binding two fractional $S=1/2$
spinons to an integer $S=1$ triplet. The size of this bound object
tends to infinity for $J_\perp\to0$.

For the value $J_\perp=J_\parallel$ needed to compare with the
square lattice it is a priori unclear which picture is superior.
Note that we are {\em not} aiming at the question which are the
elementary excitations in the strict sense of the word since this
question concerns only the lowest energies.
But we ask which excitations
describe the physics at all energies best. We show that the
approach (i) in terms of integral dressed excitations works
excellently. It provides for the first time a quantitative
description of the spectral densities on all energy scales.

Previous investigations in gapped 1D systems were also concerned
with the nature of the  excitations. For instance,
Sushkov and Kotov compared quantum antiferromagnetism to quantum
chromodynamics in the sense that the $S=1/2$ spins correspond to
quarks and integer triplets to (vector) mesons \cite{sushk98}.
Zheng et al.\ described the deconfinement of fractional spinons in
dimerized, gapped spin chains  on decreasing dimerization in terms
of integer triplet excitations \cite{zheng01b}.

We construct a mapping by a continuous unitary transformation
(CUT) \cite{wegne94} of the Hamiltonian in (\ref{hamilton}) to an
effective $H_{\rm eff}$ conserving the number of rung-triplets:
$[H_0,H_{\rm eff}]=0$. The ground state of $H_{\rm eff}$ is the
rung-triplet vacuum \cite{knett00a}. In this way a tremendously
complicated many-body problem is reduced to a tractable few-body
problem! The mapping needs an auxiliary variable $\ell$ running
between 0 and $\infty$.
 The Hamiltonian $H(\ell)$ is transformed from $\ell=0$ ($H(0)=H$) to
$\ell=\infty$ ($H(\infty)=H_{\rm eff}$) by
\begin{eqnarray}
\label{flow}
{dH}/{d\ell} &=& [\eta(\ell), H(\ell)]\\
\label{choice}
\eta_{i,j}(\ell) &=& \mbox{sign}(Q_{i,i}-Q_{j,j})
H_{i,j}(\ell)\ ,
\end{eqnarray}
where $\eta$ defines the mapping.  The matrix elements
$\eta_{i,j}$ and $H_{i,j}$ are given in an eigen-basis of  the
number of rung-triplets $Q=H_0$.
 Eqs.\ (\ref{flow},\ref{choice})
constitute a general, versatile prescription for any many-body
problem to obtain an effective model in terms of elementary
excitations, the number of which is counted by $Q$. The choice
(\ref{choice}) eliminates all parts of $H$ changing the number of
triplets while retaining a certain simplicity in $H(\ell)$ for
intermediate values of $\ell$ \cite{knett00a,mielk98}, namely that
the number of rung-triplets is changed at most by $\pm 2$. For
details on the form of $H_{\rm eff}$ the reader is referred to
Ref.\ \onlinecite{knett00a}. To determine response functions the
physical observable under study must be subject to the same
unitary transformation (\ref{flow}) as the Hamiltonian. Hence the
CUTs are based on a very clear-cut concept. Indeed, this concept
rendered the computation of bound states in higher order possible
\cite{uhrig98c,knett00b}. Conventional cluster techniques
\cite{gelfa00} allow equally to determine the energies of bound
states very accurately if they are supplemented by similarity transformations
substituting the unitary transformation \cite{trebs00,zheng01a}.
A diagrammatic approach with hard-core bosons
\cite{sushk98,kotov99} describes the excitations equally in terms of
 triplets with qualitatively similar results
deviating, however, quantitatively  for $x\gtrapprox 0.5$.

 We were able to  keep
terms in $H(\ell)$ up to order 14 in $x:=J_\parallel/J_\perp$
for the kinetic energy (triplet hopping) and up to order 13 for
the triplet-triplet interaction.
Observables were transformed up to order 10
allowing for the
first time to determine true multi-particle {\em continua}, i.e.\
spectral densities, not only spectral weights, from a
non-simulation approach.
Our perturbative
approach is supplemented by standard extrapolations and
optimizations of the resulting series. It turned out that it is
possible to determine the spectral densities within
about 5\% accuracy for $x\approx 1$. For lower values of
$x$, the accuracy improves rapidly so that the results  at
$x\lessapprox 0.6$ can be considered exact. The results are not
plagued by any finite size nor by any finite resolution effects
thereby providing for the first time predictions in great detail.

For experimental relevance \cite{windt01}
 and for comparison to the square
lattice we present results for $J_\parallel=J_\perp$.  For
orientation Fig.\ 1 depicts the dispersion of the gapped
elementary triplets \cite{barne93,greve96,shelt96}, the resulting
lower and upper edge of the 2-triplet continuum and the bound
states in the $S=0$ and the $S=1$ channel
\cite{uhrig96b,sushk98,damle98,trebs00,zheng01a,windt01}. The
spectral densities $I(\omega)$ are computed for the resolvent
\begin{equation}
I(\omega) =-{\pi}^{-1}\Im\left\langle 0\left|R
(\omega+E_0-H)^{-1}R\right|0\right\rangle_{\rm retarded}
\end{equation}
for various operators $R$ connecting the ground state to excited
states of different spin and parity. Odd (even) parity with
respect to reflection (RSc) about the centerline of the ladder
implies an odd (even) number of rung-triplets. In Fig.\ 2 the
distribution of spectral weight among the sectors of different
number of rung-triplets is depicted. For local (i.e.\
$k$-integrated) excitations the relative intensities $I^{\rm r}_n=
I_n/I_{\rm tot}$ are plotted where the intensity $I_n$ stands for
the $\omega$-integrated spectral densities $I(\omega)$ in the
sector of  $n$ rung-triplets. Recall that by our transformation
 the number of rung-triplets has become a good quantum number.
The
total intensities $I_{\rm tot}=\int_{0+}^\infty I(\omega) d\omega
=\sum_{n=1}^\infty I_n$ are accessible
 through the sum rule $I_{\rm tot}= \langle 0|R^2|0\rangle - \langle
0|R|0\rangle^2$.

For $x=0$, the excitation of a local rung-triplet exhausts the
entire spectral weight for $S=1$, i.e.\ $I^{\rm r}_1(x=0)=1$. On
increasing $x$ the triplet becomes dressed and the quasi-particle
weight decreases. The sum of all relative intensities must be
unity. Up to  $x\approx 1$ our result (dashed lines) deviates from the
sum rule less than 3\% giving evidence for the reliability of the
extrapolations.
 For larger  $x$ the weight is overestimated increasingly.
For $x=1$ the $I_1$ and the $I_2$ contributions  exhaust
74\%+19\%=93\% for $S=1$ and 77\% for $S=0$ of the total weight.
The neglect of contributions from  more triplets is hence well
justified. A description in terms of 1, 2 or 3 integer excitations
works perfectly without need to resort to spinons.

In Figs.\ 3 and 4 the $k$-resolved spectra are shown for $n=1$
(Fig.\ 3a) and $n=2$ (other panels). In both spin channels the
2-triplet spectral densities display important and fine-structured
continua bounded by square-root edges. Generally the interaction
shifts weight down to lower energies. For momenta not to far away
from $\pi$, bound states (grey lines in Figs.\ 3 and 4) leave the
continua. These bound states carry a large fraction of 2-triplet
weight.
 The differences between Figs.\ 4a and b result from the different
parity of the two excitation operators at $k=\pi$
with respect to reflection about the rung-direction (RSa), see also
Ref.\ \onlinecite{windt01}.

A very intriguing feature is the line of square root singularities
seen within the continuum in Figs.\ 3b (dark grey line) and 4b. It
increases from $k\approx\pi/3$ to $k=\pi$ being particularly
clearly visible above $k=\pi/2$. Below this line the larger part
of the spectral weight is found. The momentum dependence of these
mid-band square root singularities is strongly reminiscent of the
upper limit of the 2-spinon continuum in single chains
\cite{karba97}, to which it will evolve for $J_\perp \to0$. The
striking fact that  a 2-triplet description is capable to yield
 2-spinon features leads us to the conjecture that the important
point is {\it not} the fractionality of the excitations {\it but}
the correct description of multi-particle excitations.
Support for our conjecture is provided by the
computation of bound states of two spinons \cite{affle97,uhrig99a}
in terms of two triplets \cite{uhrig96b,zheng01b} in dimerized
spin chains.

 In Fig.\ 4 for small values of $k$
a second upper peak occurs which is due to density-of-states
effects resulting from the shallow dip that the 1-triplet
dispersion (thin solid line in Fig.\ 1) displays at $k=0$. This
local minimum in the dispersion induces a van-Hove singularity
which is smeared out by the interaction to the additional peak. It
appears that the two parts of the dispersion from $k=0$ to
$k\approx 0.4 \pi$ and from $k=\pi$ to $k\approx \pi/2$ act as
(almost) independent bands. With the help of this picture, the
rather sharp resonance at about $k=\pi/2$ in Fig.~4 is interpreted
as bound state of the part of the dispersion for $|k|\in
[0,0.4\pi] $. It is not an absolutely stable state, i.e.\ not a
$\delta$-peak of zero width, because it may still decay into
states with $|k|\in [0.4\pi,\pi]$. To be more specific, the
dispersion for small values of $k$ can be described approximately
by $\omega(k) \approx J_\perp(1.9-0.1\cos(k/0.4))$ (see Fig.\ 1)
 so that the two-particle
kinetic energy $\Omega(k,q)$ at total momentum $k$ and relative momentum $q$
given by
$\Omega(k,q)=\omega(k/2+q)+ \omega(k/2-q)$ becomes dispersionless for
$k=0.4\pi$:
$\Omega(0.4\pi,q)\approx 3.8 J_\perp$ for values of $k/2\pm q$ in the range of
the shallow dip. This implies that the relative kinetic energy is negligible
and interaction effects  dominate ! Indeed,  Fig.~4 shows
that the resonance is shifted by
attractive interaction from $3.8J_\perp$ to $\approx 3.5 J_\perp$.

Our results  predict many  features in experiment. By inelastic
neutron scattering the 2-triplet $S=1$ response (Fig.\ 3b) is
detectable if the momentum along the rungs can be tuned to zero so
that only the symmetric combination $(S^z_{1,i}+S^z_{2,i})/2$
leads to excitations. Then the 2-triplet response (Fig.~3b) is
separated clearly from the stronger 1-triplet response (Fig.~3a).
To gain further insight in the 2-triplet interaction in the $S=1$
channel we strongly suggest such experiments.

Optical spectroscopy in terms of bimagnon-plus-phonon absorption
detects a weighted superposition of the curves in Fig.\ 4
providing the first experimental verification of the existence of
the bound state in the $S=0$ sector \cite{windt01}. Raman
spectroscopy measures in leading order \cite{shast90} the curves
at $k=0$ in Fig.\ 4 \cite{sugai99}.
 The theoretical results agree very well
with experiment \cite{schmi01} though some deviations remain.
These are are most probably due to the general presence of
non-negligible cyclic exchange terms in the cuprates
\cite{schmi90,honda93,brehm99,mulle01,matsu00b,windt01}.

Our findings on the distribution of spectral weight show that
large weight at high energies does not imply the necessity to
resort to fractional excitations contrary to a frequently used
line of argument, see e.g.\ Ref.\ \onlinecite{ho01}. Integer
excitations allow to describe spectral densities of spin ladders
in great detail on all energy scales. It turns out to be essential
to describe multi-particle excitations adequately. We find also
that this task is tractable since the sectors of low
particle-number dominate clearly.

We expect that a similar
 approach with integer excitations will also allow to
describe undoped 2D cuprates quantitatively. Continuous unitary
transformations proved to be a very powerful concept to deal with
 dressed excitations. The above findings call for a
renewed discussion of the importance of fractional excitations in
low-dimensional strongly correlated electron systems such as doped
and undoped high-$T_c$ superconductors.

We acknowledge U.~L\"ow and E.~M\"uller-Hartmann for helpful
discussions, T.~Kopp and T.~Nunner for providing numerical data
prior to publication and the DFG for support in SP1073.


\begin{figure}[tb]
  \begin{center}
    \includegraphics[width=10cm]{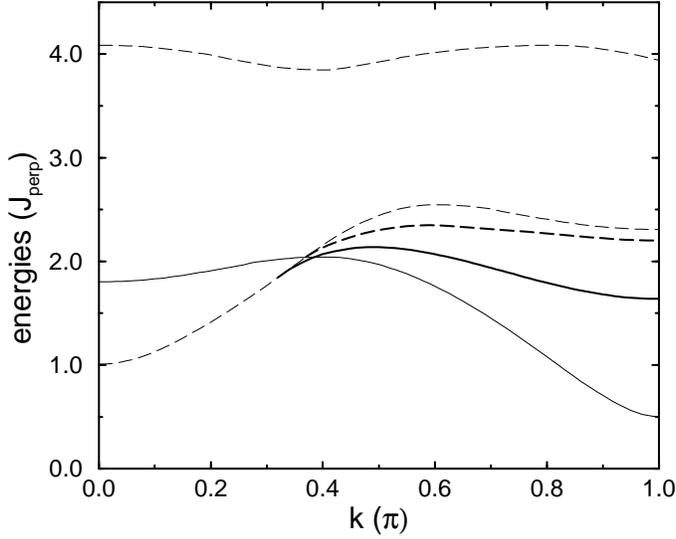}
    \caption{Dispersions at $J_\parallel=J_\perp$. Elementary triplet
      (thin solid), lower and upper 2-triplet continuum edge (thin dashed),
      bound 2-triplet states
      with $S=0$ (thick solid) and $S=1$ (thick dashed).}
  \end{center}
\end{figure}

\begin{figure}[tb]
 \begin{center}
    \includegraphics[width=9cm]{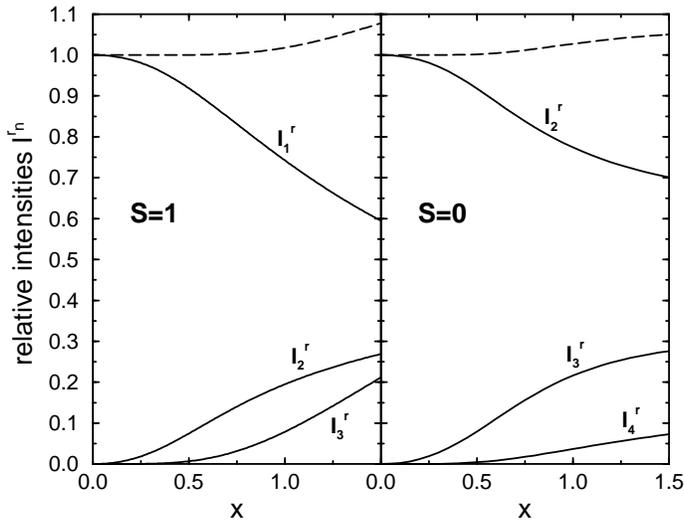}
    \caption{Relative intensities $I^{\rm r}_n$ of sub-sectors of different
      number $n$ of triplets and their sum (dashed) for local excitations
      $R_i$. Left panel $S=1$: $R_i={\bf S}^z_{1,i}$; right panel $S=0$:
      $R_i={\bf S}_{1,i}{\bf S}_{1,i+1}$.}
  \end{center}
\end{figure}

\clearpage

\begin{figure}[b]
 \begin{center}
    \includegraphics[width=12cm]{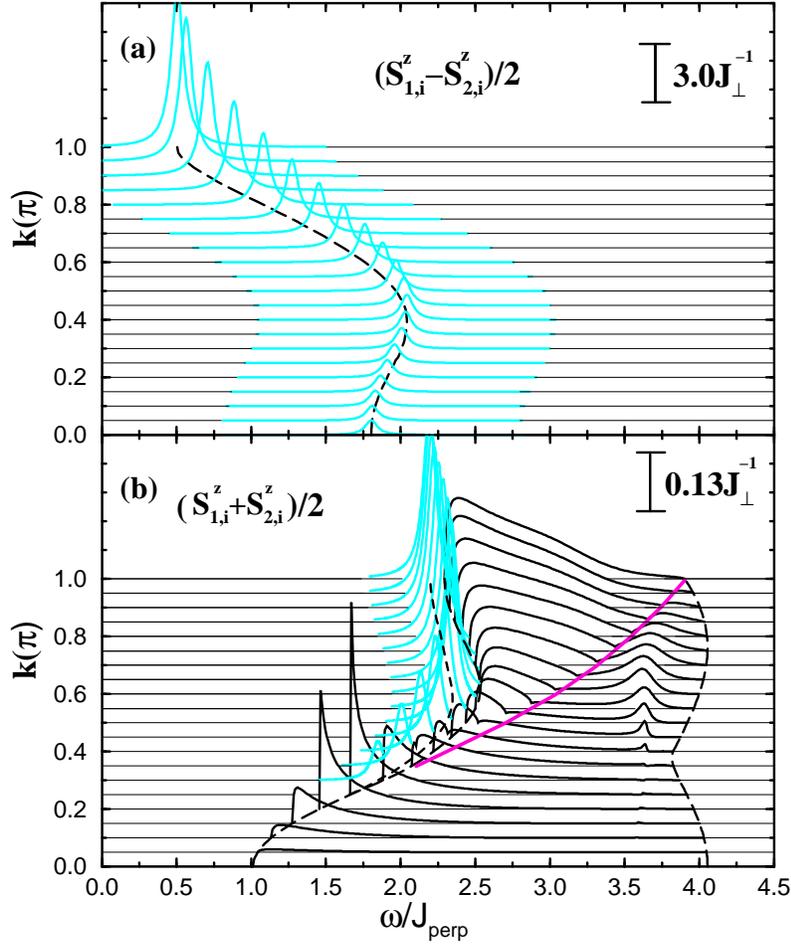}
    \caption{$k$-resolved spectral densities for operators indicated
      ($S=1$);
    $\delta$-peaks (grey lines) broadened for visualization by
    $J_\perp / 20$ and inserted in front of the unbroadened continua. The bound
    state energies and the continuum edges are also shown (dashed lines).
    (a)  1-triplet peak ; (b) 2-triplet continuum (multiplied by 4 to show details) and bound state.
    The dark grey line is a guide to the eye linking
    the mid-band square root singularities.}
  \end{center}
\end{figure}

\clearpage

\begin{figure}[tb]
 \begin{center}
    \includegraphics[width=12cm]{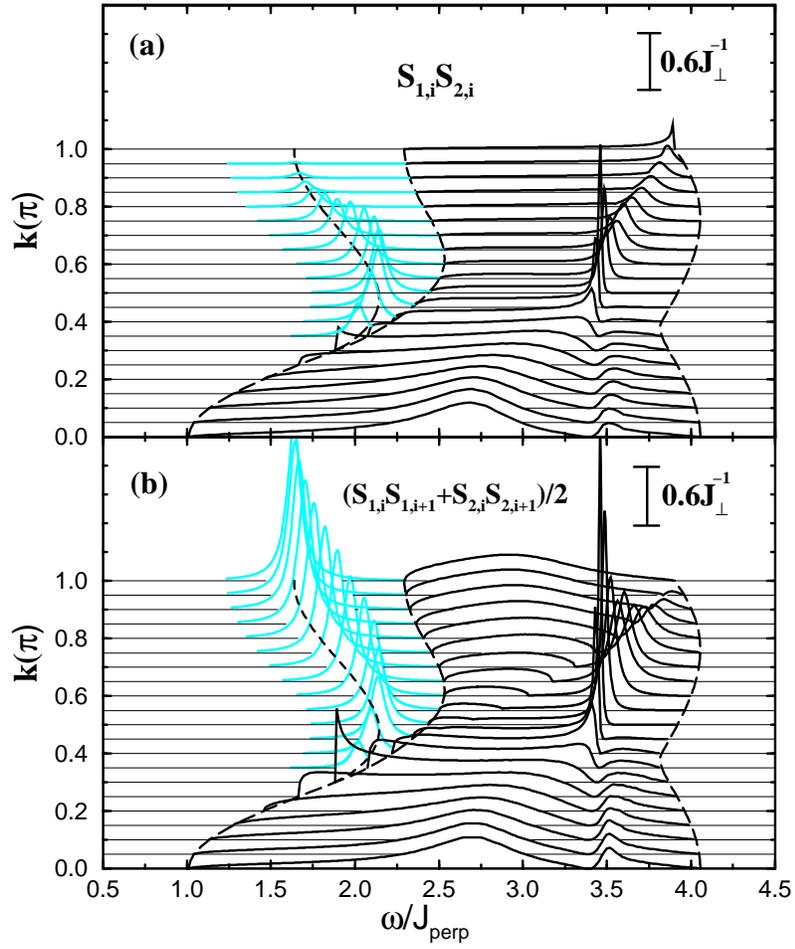}
    \caption{Same as in Fig.\ 3 for $S=0$; (a,b) 2-triplet continuum
      and bound state. In (b) continuum multiplied by 4 to show details.}
  \end{center}
\end{figure}

\end{document}